# Simplifying Integration of Custom Controllers in Exergames

**Hassan Ali Khan**
NC State University
Raleigh, US
**hakhan@ncsu.edu**

**Muhammad Asbar Javed**
University of Lahore
Lahore, Pakistan
asbarjavaid@gmail.com

**Amnah Khan**
LUMS
Lahore, Pakistan
1630043@lums.edu.pk

**ABSTRACT**
Despite of the established evidence in favor of exergames for physical rehabilitation their use is limited in Pakistan. In our user study with game developers (N=62), majority (67.7%) of the participants believed that exergames' popularity will increase if cheap alternatives of body tracking devices are available. Perhaps, custom controllers can be used as an affordable alternate input source in exergames but the lack of hardware programming knowledge and shortage of experience in the embedded programming attribute to the little involvement of game developers (11.3% of the participants) in the area of exergames. This paper presents a library for the integration of Arduino based (open-source and low-cost) tailored controllers to be used as input source in Unity3D (most preferred game development engine by 88.7% participants) based exergames. The interface to the library proposes a flexible and easy structure for programming and serve as a template application for a range of exergames.

**ACM Classification Keywords**
H.5.m. Information interfaces and presentation (e.g., HCI): Miscellaneous; See http://acm.org/about/class/1998 for the full list of ACM classifiers. This section is required.

**INTRODUCTION**
The clinical importance of Motion based Games for Health (MGH - video games that are also a form of exercise) is growing for rehabilitation and MGH or exergames are increasingly being adopted as a major tool in the rehabilitation plans for various neuromuscular disorders [1,2,3]. Despite their proven benefits of adherence to exercises, enjoyment and motivation [1], the MGH or exergames are still not popular in the developing countries such as Pakistan.

Underscoring the difference in exergames research in the world and Pakistan, we mined papers from ACM digital library and IEEE Xplore digital library, using the keywords, 'exergames' and 'motion-based health games' from March 2014 to March 2018. We found that number of research articles published from Pakistan are very low (57) as compared to overall exergames related publications (4,454) in the period of four years. For a country with shortage of medical professionals, there could be a huge market for culturally relevant and affordable exergames.

To get the professionals' opinion on popularity of exergames, we conducted an open-ended survey with 62 game-developers (28 months average game development experience) from Pakistan. According to the survey results, 80.6% game developers have heard of exergames but only 11.3% of them had developed any. HTC vive is most famous body tracking device among developers (mentioned by 72.5% of the participants), followed by Oculus Rift (64.5%) and Kinect (59.7%). It was further highlighted in the survey analysis that the price range of body tracking solutions (HTC VIVE, MS Kinect and Oculus Rift) which act as input source for exergames are significantly higher than the buying capacity of the intended audience. 67.7% of the game developers believed that exergames popularity will increase if cheap and affordable alternatives of body tracking devices are available. Regarding game development engine, most of the participants (87%) mentioned that Unity3D is their first preference for the game development and it was also revealed that Arduino is the most familiar embedded device/prototyping board among (~63%) participants of the survey. Participants with notable majority (72.6%) also mentioned that they are not comfortable with low level programming due to the unfamiliarity with hardware programming, low level programming complexity and lack of training in the embedded programming. However, 88.7% game developers agreed that they can use embedded devices or custom controllers for exergames if they don't have to program it themselves.

As a resolution to the above expressed professionals' concerns regarding the popularity of exergames, we present an extendible and generic library for the integration of custom (low cost and open-source Arduino based) user tracking controllers as an input source of exergames developed in Unity3D.

**RELATED WORK**

Considering the output of the survey and current trends of research involving smart embedded devices (specifically Arduino) as the source of input controller for Exergames, we searched the literature regarding the general-purpose libraries offering communication between Unity3D based games and Arduino boards (integrated with sensors and input-output modules) that can serve as a template application for any project to optimize the development time and cost.

There are number of open-source Unity-Arduino communication tools but they are inadequate in usability. Uniduino [4] only exposes methods for serial communication between Arduino and Unity3D. ARDUnity[5] can generate and burn scratch program file for Arduino dynamically from Unity3D editor but it hinders the utilization of Unity platform's event-driven approach. Arduino-to-Unity [6], Arduino-Unity-Bridge [7] and unity-Arduino-serial [8] have a very limited use as they are developed specifically for one application/game and will need the game developer to modify code files for both arduino and unity platforms. SerialFinder_Arduino [9] and SerialCommUnity [10] expose only the basic functions of Arduino as digitalRead, digitalWrite, analogRead and analogWrite in Unity 3D and require the user to code on both Unity and Arduino for extra input sensors.

In order to address the above-mentioned shortcomings, we have developed an open-source Tailoring the Input Source for Exergames (TISE) library for the Arduino-Unity3D communication. TISE can help the game programmers in developing a custom controller based exergame completely in Unity3D, without having to code in Arduino IDE

**TISE IMPLEMENTATION**

TISE library consists of template *Arduino scratch program* and a *utility library* (targeting .Net standards 2.0, that allow library functions to be called by any .Net application). This dynamically linked library (DLL) serves as the Unity plugin as it is possible to compile a script that uses functions of DLL using Mono compiler in Unity3D, which supports .Net framework runtime. TISE library uses serial communication for the transmission of messages between Arduino and Unity3D application. Both scratch program and unity plugin use read and write buffers for serial communication.

The communication is initiated by a handshake in which application of Unity3D sends a control signal to the Arduino board and the board acknowledges the message (Figure 1). The messages sent by Unity3D application are first parsed in Arduino and corresponding routine is initiated as per the switch statements (Figure 1). The user will have to burn the template application on Arduino to set or get the values of sensors and input-output modules in the Unity3D editor and trigger some in-game actions e.g. making a game character

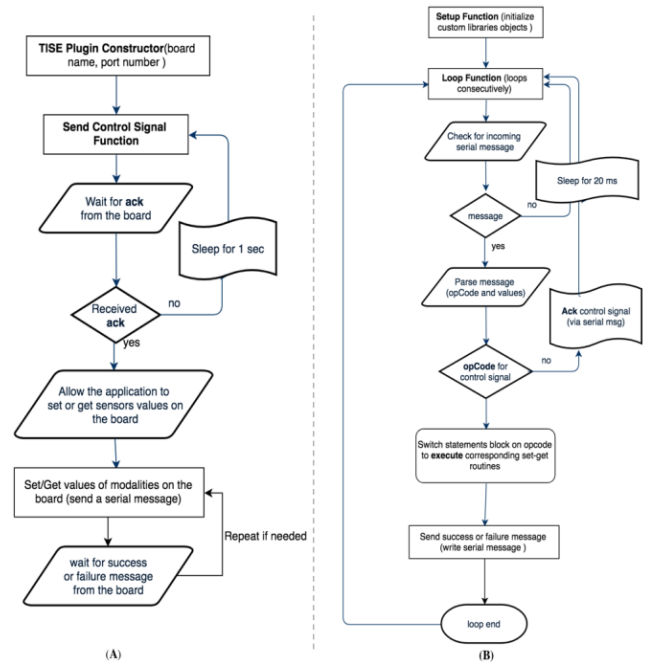

**Figure 1. (A)** Flow chart for the execution of .Net library (Unity3D Plugin), **(B)** Flow chart for the execution of TISE Arduino template application.

jump off the hurdle depending on a certain value of the gyroscope sensor.

Currently, fifteen sensors and input-output modules (Push Button, 16x2 and 16x4 LCD Module, LM35 Sensor, Ultrasonic Sensor HC-SR04, Light Dependent Resistor, TowerPro SG90 servo motor, DC Motor, 5V IR LED, TSOP382 IR Receiver, MQ-X Gas Sensor, Digit Seven Segment, Potentiometer, 5mm LED, Microphone sound sensor, SW-420 Motion Vibration Sensor, and YL-44 Buzzer Module) can be modulated in TISE library completely through the Unity3D editor.

**CONCLUSION AND FUTURE WORK**

TISE library provides flexible integration of different physical sensors in Unity3D projects and decouples the requirement of low-level programming knowledge from the use of custom controller in exergames. It can help the game programmers in developing immersive and interactive exergames without having to worry about programming the embedded devices and minimizing the game development time. It can also shift the focus of exergames from balance training (Coordination exercises) to muscle strengthening (isotonic exercises) which is currently lacking in most of the rehabilitative exergames.

Our ongoing work involves support for more sensors in the library. We have also planned to develop the cross-platform C++ version of TISE library to support the native mobile applications environments (Android using NDK & iOS) to facilitate the development of affordable, fun and ubiquitous exergames.